\documentclass[twocolumn,english,prl]{revtex4-1}
\usepackage[T1]{fontenc}
\usepackage[latin9]{inputenc}
\usepackage{color}
\usepackage{float}
\usepackage{textcomp}
\usepackage{amstext}
\usepackage{amssymb}
\usepackage{graphicx}
\usepackage{esint}

\makeatletter

\usepackage{float}
\usepackage{babel}

\usepackage{babel}

\usepackage{babel}

\usepackage{babel}

\makeatother

\usepackage{babel}
\begin{document}

\title{Linking microscopic and macroscopic response in disordered solids}

\author{Daniel Hexner}
\email{danielhe2@uchicago.edu}

\affiliation{The James Franck Institute and Department of Physics, The University
of Chicago, Chicago, IL 60637, USA and Department of Physics and Astronomy,
The University of Pennsylvania, Philadelphia, PA, 19104, USA}

\author{Andrea J. Liu}

\affiliation{Department of Physics and Astronomy, The University of Pennsylvania,
Philadelphia, PA, 19104, USA}

\author{Sidney R. Nagel}

\affiliation{The James Franck and Enrico Fermi Institutes and The Department of
Physics, The University of Chicago, Chicago, IL 60637, USA}
\begin{abstract}
The modulus of a rigid network of harmonic springs depends on the
sum of the energies in each of the bonds due to the applied distortion:
compression in the case of the bulk modulus, $B$, or shear in the
case of the shear modulus, $\mathcal{G}$. The distortion need not
be global and we introduce a \textit{local} modulus, $L_{i}$, associated
with changing the equilibrium length of a single bond, $i$, in the
network. We show that $L_{i}$ is useful for understanding many aspects
of the mechanical response of the entire system. For example, it allows
an understanding, and efficient computation, of how each bond in a
network contributes to global properties such as $B$ and $\mathcal{G}$
and sheds light on how a particular bond's contribution to one modulus
is, or is not, correlated with its contribution to another. 
\end{abstract}
\maketitle
In rigid networks, nodes are connected by struts. In the simplest
case, the connections can be approximated by unstretched central-force
harmonic springs. Such networks have been demonstrated to be extraordinarily
tunable in their mechanical properties. In particular, the ratio of
the shear modulus $\mathcal{G}$ to the bulk modulus $B$ can be varied
over its entire range, $0\le\mathcal{G}/B\le\infty$, by selectively
removing only a tiny fraction of the bonds~\cite{tuning_by_pruning}.

A global modulus $M$, such as $B$ or $\mathcal{G}$, can be expressed
as a sum of contributions $M_{i}$ from individual bonds $i$: $M=\sum_{i}M_{i}$.
Here we calculate $\Delta M_{i}$, the change in $M$ due to the removal
of bond $i$, and show within linear response that the quantities
$M_{i}$ and $\Delta M_{i}$ are related via a \emph{local} modulus,
$L_{i}$. This provides a basis for understanding the sensitivity
of $\mathcal{G}/B$ to selective bond removal.

\textit{The local modulus and its connection to bond removal:} We
consider an infinitesimal perturbation that alters 
the equilibrium length of a single spring $i$ in the network by an
amount $\delta\ell_{i}$. This local strain perturbation leads to
stresses $t_{j}$ on all remaining springs $j$. The total energy
is the sum of all the bond energies: $E=\frac{1}{2}\sum_{j}t_{j}^{2}/k_{j}$
where $k_{j}$ is the spring constant of bond $j$. We define the
local modulus as $L_{i}\equiv2E/\delta\ell_{i}^{2}$. 
The dependence on the network structure is captured in $S_{i}^{2}\equiv\frac{L_{i}}{k_{i}}$.

We now show that $\Delta M_{i}=M_{i}/S_{i}^{2}$. For simplicity,
we consider all springs to have the same spring constant, $k$. (The
Supplementary Information shows that this exact relation is also true
in the general case with different $k_{i}$.) The derivation is based
on the formalism of states of self stress \cite{CALLADINE,Maxwell},
which are the set of combinations of tensions $\left\{ t_{i}\right\} $
resulting in force balance on all nodes. We use $\alpha$ and $\beta$
to index these states of self stress: $s_{\alpha,i}$ is the tension
on bond $i$ in the state $s_{\alpha}$. There are an extensive number
of such combinations and it is useful to define an orthonormal basis
satisfying $s_{\alpha}\cdot s_{\beta}=\delta_{\alpha,\beta}$.

Within linear response the energy of a deformation, derived in \cite{PELLEGRINO2,PELLEGRINO1,Lubensky_rev,WyartReview},
can be expressed in terms of $s_{\alpha}$. The resulting energy and
bond tensions are given by: 
\begin{equation}
E=\frac{k}{2}\sum_{\alpha=1}^{N_{s}}\left(s_{\alpha}\cdot e_{M}\right)^{2},\label{eq:f_el-1}
\end{equation}
\begin{equation}
t_{i}=k\sum_{\alpha=1}^{N_{s}}\left(s_{\alpha}\cdot e_{M}\right)s_{\alpha,i}.\label{eq:tensions}
\end{equation}
where $N_{s}$ is the number of states of self stress and $e_{Mi}=\sum_{\mu\nu}\delta r_{i,\mu}\epsilon_{M\mu\nu}\hat{\delta r_{i,\nu}}$
are the affine bond extensions, which depend on the strain tensor
$\epsilon_{M\mu\nu}$ corresponding to the applied deformation for
modulus $M$ and the bond vector $\delta r_{i}$. Here $\mu$ and
$\nu$ index the spatial coordinates.

There is large degeneracy in choosing a basis since any linear combination
of $s_{\alpha}$ is also a state of self stress. For convenience we
choose $s_{1}$ to be in the direction of $e_{M}$ so that $e_{M}\cdot s_{\alpha\ne1}=0$
and only the $\alpha=1$ term contributes to the deformation energy
\footnote{To find $s_{1}$ we assume an arbitrary basis $s_{\alpha}*$ and then
$s_{1}=C\sum_{\alpha}\left(e\cdot s_{\alpha}*\right)s_{\alpha}*$
where $C$ is a normalization constant. This satisfies by definition
$e\cdot s_{\alpha}=0$ for all $\alpha>1$. }. With this choice of basis, 
\begin{equation}
M_{i}=Ms_{1,i}^{2}.\label{eq:modulus}
\end{equation}
where the modulus $M=\frac{k}{V}(e_{M}\cdot s_{1})^{2}$ and $V$
is the volume.

In order to compute $\Delta M$, the change in a modulus after bond
$i$ is removed, we need a new basis in the pruned network, $s'_{\alpha}$.
Fortunately, this can be expressed using $s_{\alpha}$ of the unpruned
network. To see this, we note that any linear combination of $s_{\alpha}$
which has zero tension on bond $i$ is also a state of self stress
of the pruned network, since force balance is still obeyed.

We introduce a new state of self stress that depends on $i$, the
targeted bond: $S_{i}=\sum_{\alpha=1}^{N_{s}}s_{\alpha,i}s_{\alpha}$.
$S_{i}$ is independent of the choice of basis and has several nice
properties. First, using Eq. \ref{eq:tensions} with $e_{Mj}=e_{j}=\delta_{ij}$,
one can verify that the stress on bond $j$ resulting from a unit
change in equilibrium length of bond $i$ is $t_{j}=k\left[S_{i}\right]_{j}=k\left[S_{j}\right]_{i}$\footnote{$\left[S_{i}\right]_{j}$ is related to the dipole response studied
in \cite{Jamming_SSS,dipole_response} and decays as a function of
distance between bonds with a characteristic length scale that diverges
as $\Delta Z\rightarrow0$. }. We can compute the energy from these values of the stresses, $E=\frac{1}{2}k\sum_{j}\left[S_{i}\right]_{j}^{2}$.
We also compute the energy from Eq. \ref{eq:f_el-1} using $e_{j}=\delta_{ij}$\footnote{The equality can be verified by noting that $\left[S_{i}\right]_{j}$
is a projection matrix such that $\sum_{j}\left[S_{i}\right]_{j}\left[S_{j}\right]_{k}=\left[S_{i}\right]_{k}$.}: 
\begin{equation}
E=\frac{k}{2}\sum_{\alpha=1}^{N_{s}}s_{\alpha,i}^{2}=\frac{k}{2}\left[S_{i}\right]_{i}.\label{eq:eq2}
\end{equation}
This implies that $L_{i}=k\left[S_{i}\right]_{i}=kS_{i}^{2}\equiv kS_{i}\cdot S_{i}$
is the local modulus.

Using $S_{i}^{2}$ we now prove that 
\begin{equation}
s'_{1}=C\left(s_{1}-\frac{s_{1,i}}{S_{i}^{2}}S_{i}\right)\label{eq:sigma_1-1}
\end{equation}
(where $C=\left(1-s_{1,i}^{2}/S_{i}^{2}\right)^{-1/2}$ is a normalization
constant) is the only state of self stress that contributes to the
modulus $M$ after bond $i$ is removed. To this end it must be shown
that the remaining, $s'_{\alpha>1}$, orthogonal to $s'_{1}$, are
orthogonal to the applied strain, $e_{M}$. The vector space orthogonal
to $s'_{1}$ can be constructed explicitly from the $N_{s}-1$ states
of self stress $s_{\alpha>1}$ using linear combinations of $\tilde{s}_{\alpha}=s_{\gamma,i}s_{\alpha}-s_{\alpha,i}s_{\gamma}$,where
both $\alpha>1$ and $\gamma>1$. This is constructed so that the
tension on the targeted bond $i$ is zero. To see that these are orthogonal
to $s'_{1}$, we note that by the definition $s_{1}\cdot\tilde{s_{\alpha}}=0$,
and that $S_{i}\cdot\tilde{s}_{\alpha}=0$ as can be verified using
the definition of $S_{i}$. Since the basis vectors $s'_{\alpha>1}$
are all linear combinations of $s_{\alpha>1}$, which were chosen
to be orthogonal to $e_{M}$, also $s'_{\alpha}\cdot e_{M}=0$ thus
completing the proof.

Using Eq.~\ref{eq:f_el-1} and noting that within linear response
the modulus is proportional to elastic energy, the modulus after bond
$i$ is removed is therefore: $M'=\frac{k}{V}(e_{M}\cdot s'_{1})^{2}$.
The change in modulus, $\Delta M\equiv M-M'$ due to the removal of
bond $i$, is 
\begin{equation}
\Delta M_{i}=M_{i}/S_{i}^{2}.\label{eq:DM-1}
\end{equation}
This is the central equation on which our subsequent analysis is based.

\textit{Networks created from jammed configurations:} Our numerical
results are based on networks derived from jammed packings of particles
\textendash{} a ubiquitous model for amorphous materials \cite{Durian1995}.
Configurations are prepared by standard methods\cite{Review,angle_avg};
soft frictionless repulsive spheres are distributed randomly in space
and the system's energy is minimized to produce a jammed configuration
in which the coordination number, $Z$ exceeds the minimum required
for stability, $Z_{iso}$~\cite{Goodrichfinitesize}.

The system is then converted into a spring network by replacing the
spheres with springs connecting the centers of interacting particles.
We 
remove any stresses by setting the equilibrium spring length to the
inter-particle distance. Such networks capture many of the key properties
of jammed packings~\cite{Silbert2006}, such as the scalings of the
bulk and shear moduli~\cite{Ellenbroek2009}. 
We characterize these networks by their excess coordination, $\Delta Z\equiv(Z-Z_{iso})$.


\textit{Relation of $M_{i}$ to $S_{i}^{2}$:} We focus on two different
global moduli $M$, the compression modulus $B$ and the modulus corresponding
to simple shear in the $xy$-direction, $\mathcal{G}\equiv C_{xyxy}$;
our results hold for other shear elastic constants, such as pure shear
$\frac{1}{4}\left(C_{xxxx}+C_{yyyy}\text{\textminus}2C_{xxyy}\right)$,
as well. Equation \ref{eq:DM-1}, together with the condition that
the generalized modulus is non-negative after bond removal, implies
$M_{i}/M<S_{i}^{2}$. 
Both $B_{i}$ and $\mathcal{G}_{i}$ are strongly correlated with
$S_{i}^{2}$. Figure~1 plots the conditional average of $B_{i}$
and $\mathcal{G}_{i}$ for a given value of $S_{i}^{2}$, denoted
as $\left\langle B_{i}\left(S_{i}^{2}\right)\right\rangle $ and $\left\langle \mathcal{G}_{i}\left(S_{i}^{2}\right)\right\rangle $.
Excluding the largest values of $S_{i}^{2}$, both $\left\langle B_{i}\left(S_{i}^{2}\right)\right\rangle $
and $\left\langle \mathcal{G}_{i}\left(S_{i}^{2}\right)\right\rangle $
are proportional to $S_{i}^{2}$. A plateau in $\left\langle B_{i}\left(S_{i}^{2}\right)\right\rangle $
exists at large $S_{i}^{2}$ that is more prominent in $d=2$ 
(see Supplementary Information) The insets to Fig.~1 show that both
$\left\langle \Delta B_{i}\left(S_{i}^{2}\right)\right\rangle $ and
$\left\langle \Delta\mathcal{G}_{i}\left(S_{i}^{2}\right)\right\rangle $
are nearly completely uncorrelated with $S_{i}^{2}$.

\begin{figure}
\includegraphics[scale=0.35]{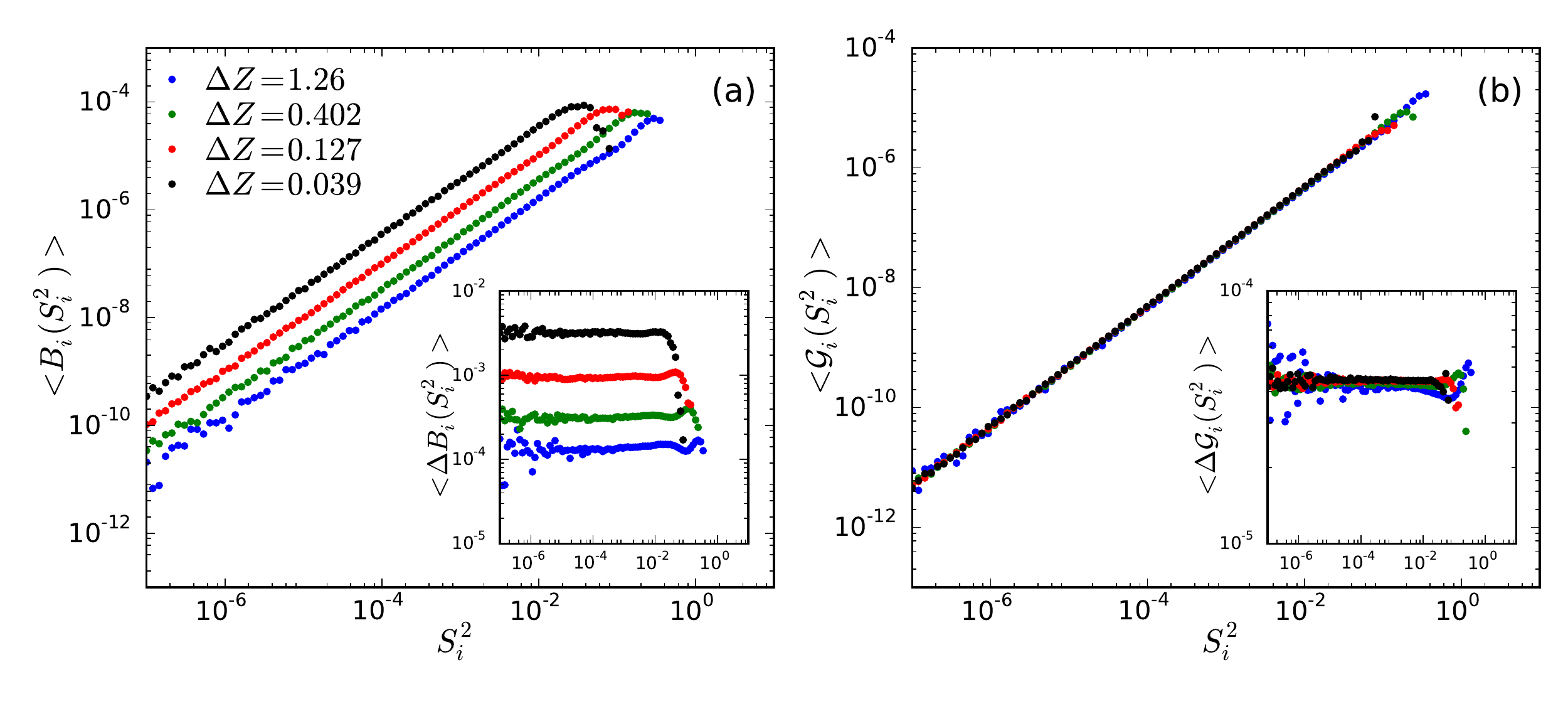} 

\caption{a) The conditional average $\left\langle B_{i}\left(S_{i}^{2}\right)\right\rangle $
and b) $\left\langle \mathcal{G}_{i}\left(S_{i}^{2}\right)\right\rangle $
are proportional to $S_{i}^{2}$ over a broad range of $S_{i}^{2}$.
In the inset the corresponding $\left\langle \Delta B_{i}\left(S_{i}^{2}\right)\right\rangle $
and $\left\langle \Delta\mathcal{G}_{i}\left(S_{i}^{2}\right)\right\rangle $
are shown and to good approximation are independent of $S_{i}^{2}$.
Data from 3d system with $N=4096$ particles. See supplementary material
at {[}URL will be inserted by publisher{]} for 2d figure. }
\end{figure}

Varying $\Delta Z$ does not change $\left\langle \mathcal{G}_{i}\left(S_{i}^{2}\right)\right\rangle $,
but does change the overall magnitude of $\left\langle B_{i}\left(S_{i}^{2}\right)\right\rangle $,
which scales as $\Delta Z^{-\lambda}$. In $d=2$, $\lambda_{2D}\approx1.4$
while in $d=3$, $\lambda_{3D}\approx1.0$. The average values of
the modulus can be related to the conditional average. In $d=3$:
$\left\langle B_{i}\right\rangle =\int dS_{i}\left\langle B_{i}\left(S_{i}^{2}\right)\right\rangle P\left(S_{i}^{2}\right)\propto\Delta Z^{-\lambda_{3D}}\left\langle S_{i}^{2}\right\rangle $
(where we substituted in the linear dependence of $\left\langle B_{i}\left(S_{i}^{2}\right)\right\rangle $)
and $\left\langle \mathcal{G}_{i}\right\rangle \propto\int dS_{i}\left\langle \mathcal{G}_{i}\left(S_{i}^{2}\right)\right\rangle P\left(S_{i}^{2}\right)=\left\langle S_{i}^{2}\right\rangle $.
Since as $\Delta Z\rightarrow0$, $\left\langle B_{i}\right\rangle \rightarrow const$
and $\left\langle \mathcal{G}_{i}\right\rangle \propto\Delta Z$,
then $\left\langle S_{i}^{2}\right\rangle \propto\Delta Z^{\lambda_{3D}}\propto\Delta Z$.
Therefore $\lambda_{3D}\approx1$, as also argued in Ref. \cite{WyartReview}
for any dimension. This analysis fails in $d=2$ because of the plateau
at high $S_{i}^{2}$ in $\left\langle B_{i}\left(S_{i}^{2}\right)\right\rangle $. 

The relation $\left\langle \mathcal{G}_{i}\left(S_{i}^{2}\right)\right\rangle \propto S_{i}^{2}$
can be understood as follows. Note that $\mathcal{G}_{i}=\frac{1}{2Vk}\left(t_{\mathcal{G}i}\right)^{2}$,
where $t_{\mathcal{G}i}$ is the tension in a bond for a shear deformation
and $V$ is the volume. Using Eq. 2: $t_{\mathcal{G}i}=k\sum_{j}e_{\mathcal{G}j}[S_{i}]_{j}$.
For a simple shear deformation $e_{\mathcal{G}i}=\epsilon\left|\delta r_{i}\right|sin\left(2\theta_{i}\right)$
where $\epsilon$ is the magnitude of the deformation, $\left|\delta r_{i}\right|$
is the length of the bond and $\theta_{i}$ is the bond angle with
respect to the y-axis. If $\theta_{i}$ have only delta-function spatial
correlations then $e_{\mathcal{G}i}$ can be considered uncorrelated
random variables, with zero mean due to isotropy. The inset to Fig.~1(b)
shows that this is a good assumption. Lastly, we assume$[S_{i}]_{j}$
is not coupled to the value of a single $e_{\mathcal{G}i}$, and depends
on the overall structure of the system so that the average is computed
only over $e_{\mathcal{G}i}$ and $[S_{i}]_{j}$ is considered constant.
Hence, $\left\langle \mathcal{G}_{i}\left(S_{i}^{2}\right)\right\rangle =\frac{1}{2Vk}\left\langle e_{\mathcal{G}i}^{2}\right\rangle \sum_{j}[S_{i}]_{j}^{2}$,
leading to 
\begin{equation}
\left\langle \mathcal{G}_{i}\left(S_{i}^{2}\right)\right\rangle =\frac{k}{2V}\left\langle e_{\mathcal{G}i}^{2}\right\rangle S_{i}^{2}.\label{eq:GSi}
\end{equation}
This approximation not only captures the dependence on $S_{i}^{2}$
but also predicts no additional dependence on $\Delta Z$ as found
for $\Delta B_{i}$. The derivation of Eq.~\ref{eq:GSi} was based
on the properties of $[S_{i}]_{j}$, the short-ranged bond angle correlations
and isotropy. As a result, we expect this relation to hold quite generally
for disordered networks.

\textit{Correlations between $B_{i}$ and $\mathcal{G}_{i}$:} In
Ref.~\cite{tuning_by_pruning} it was argued that precise control
over the ratio $\mathcal{G}/B$ required independence of bond-level
response. 
We have already shown in Fig.~1 that $\mathcal{G}_{i}$ and $B_{i}$
are both strongly correlated with $S_{i}^{2}$ and are therefore correlated
with each other. However, our analysis shows that precise control
over $\mathcal{G}/B$ depends not on $B_{i}$ and $\mathcal{G}_{i}$,
but on $\Delta B_{i}=B_{i}/S_{i}^{2}$ and $\Delta\mathcal{G}_{i}=\mathcal{G}_{i}/S_{i}^{2}$.
Indeed, we find that the values of $\Delta B_{i}$ and $\Delta\mathcal{G}_{i}$
are virtually uncorrelated with one another. We demonstrate this in
Figs.~2(a) and (b). Fig.~2(a) shows that $P\left(B_{i}\right)$
depends on the range of $\mathcal{G}_{i}$ being considered, in agreement
with Ref.~\cite{tuning_by_pruning}. By contrast, the distribution,
$P\left(\Delta B_{i}\right)$, is independent of the range of $\Delta\mathcal{G}_{i}$,
within numerical uncertainty. This implies very small correlations
between $\Delta B_{i}$ and $\Delta\mathcal{G}_{i}$.

\begin{figure}
\includegraphics[scale=0.43]{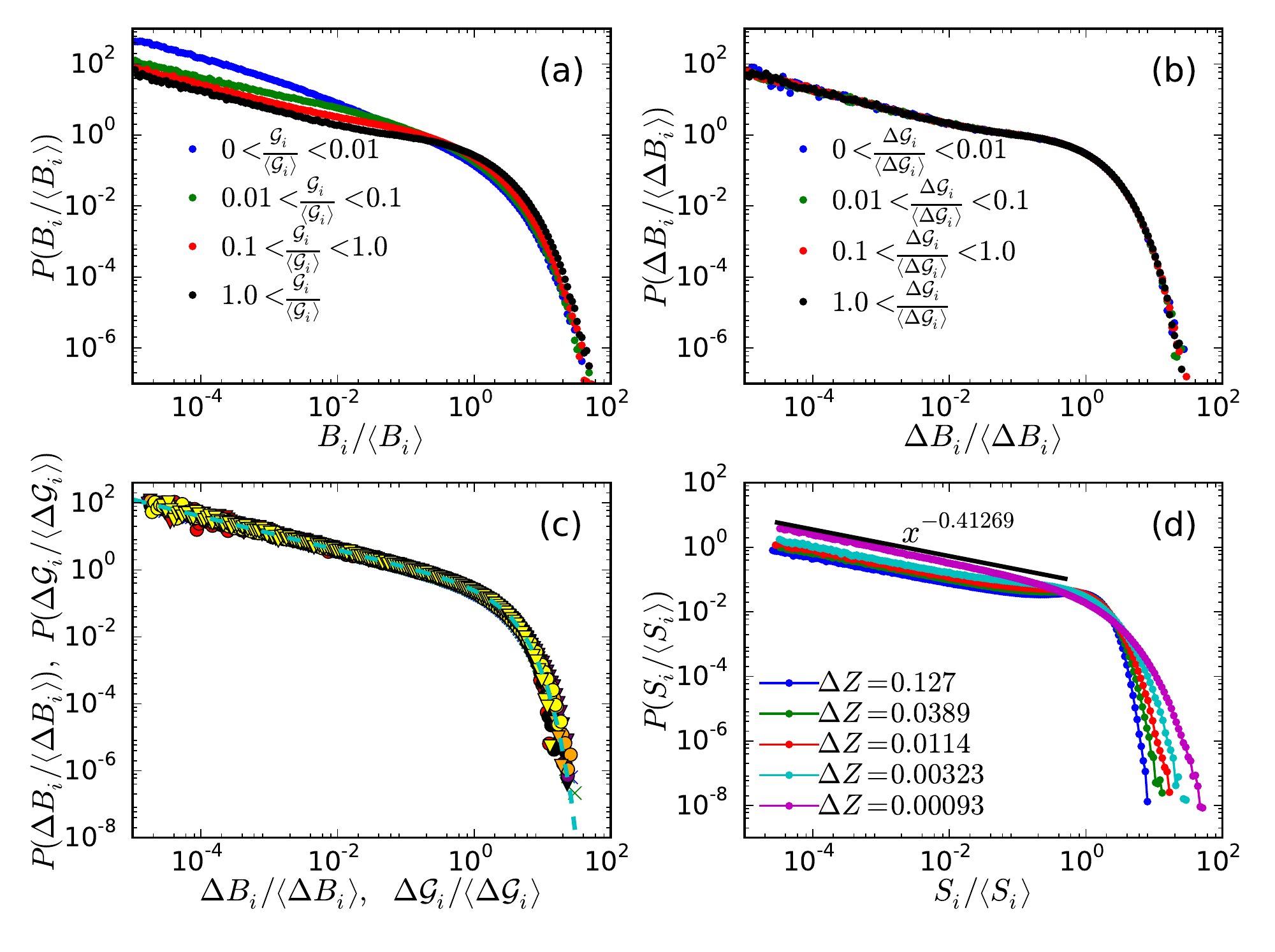}

\caption{a) $P\left(B_{i}\right)$ for different ranges of $\mathcal{G}_{i}$
and b) $P\left(\Delta B_{i}\right)$ for different range of $\Delta\mathcal{G}_{i}$
at $\Delta Z=0.127$. The collapse indicates almost vanishing correlations.
c) The universal distribution of $P\left(\Delta\mathcal{G}_{i}\right)$
for the pruned and unpruned case and the distribution of $P\left(\Delta B_{i}\right)$
for the pruned case. The cyan dashed line is the theoretical prediction
$P\left(y\right)=\frac{1}{\sqrt{2\pi}}y^{-\frac{1}{2}}e^{-\frac{y}{2}}$.
The distribution of $P\left(\Delta\mathcal{G}_{i}\right)$ in the
unpruned jammed network $\times$ blue $2d$, $\times$ green $3d$.
The following pruning protocols lead to the same distribution where
$P\left(\Delta B_{i}\right)$ is denoted by $\bullet$ and $P\left(\Delta\mathcal{G}_{i}\right)$
is denoted by $\blacktriangledown$. The the number of removed bonds
and dimensionality is also stated: red $max\Delta B_{i},\,2d,\,50$,
black $max\Delta B_{i},\,3D\,30$, purple $random,\,40,\,2d$ , orange
$min\Delta\mathcal{G}_{i},\,40,\,2d$ and yellow $max\Delta\mathcal{G}_{i},\,60\,2d$.
d) $P\left(S_{i}^{2}\right)$ for different values of $\Delta Z$.
Data from 3d system with $N=4096$ particles. See supplementary material
at {[}URL will be inserted by publisher{]} for 2D figure. }
\end{figure}

In order to quantify the correlation between $B_{i}$ and $\mathcal{G}_{i}$,
Ref.~\cite{tuning_by_pruning} used the Pearson correlation function,
$r=\frac{\left\langle B_{i}\mathcal{G}_{i}\right\rangle -\left\langle B_{i}\right\rangle \left\langle \mathcal{G}_{i}\right\rangle }{\sigma_{B_{i}}\sigma_{\mathcal{G}_{i}}}$
(where $\sigma$ denotes the standard deviation and $\left\langle ...\right\rangle $
denotes the average) and found $r\approx0.171$ in $d=2$ and $r\approx0.325$
in $d=3$. In comparison, we find that for $\Delta B_{i}$ and $\Delta\mathcal{G}_{i}$,
the corresponding Pearson correlation is $r<0.05$ in $d=2$ and $r<0.01$
in $d=3$. There is much less correlation than for $B_{i}$ and $\mathcal{G}_{i}$
\footnote{We note that small systems posses finite-size effects; in the extreme
limit of $N(Z-Z_{iso})=1$ where $N$ is the number of nodes in the
network all deformations are necessarily correlated. The correlations
quoted above are for systems large enough so that $N(Z-Z_{iso})>>1$).}. The significant correlation between $\mathcal{G}_{i}$ and $B_{i}$
exists because both quantities are correlated with $S_{i}^{2}$.

Unlike perfect lattices, jammed systems are heterogeneous such that
different bonds contribute differently to rigidity as characterized
by $S_{i}^{2}$. This produces correlations between shear and compression
since rigid regions, with a large average $S_{i}^{2}$, typically
carry more stresses regardless of the type of deformation. Our results
suggest that the correlation between $B_{i}$ and $\mathcal{G}_{i}$
can be estimated by assuming that $B_{i}=b_{i}\Delta Z^{-\lambda}S_{i}^{2}$
and $\mathcal{G}_{i}=g_{i}S_{i}^{2}$, where $b_{i}$ and $g_{i}$
are uncorrelated random variables. This assumption leads to $\left\langle B_{i}\mathcal{G}_{i}\right\rangle -\left\langle B_{i}\right\rangle \left\langle \mathcal{G}_{i}\right\rangle \propto\Delta Z^{-\lambda}\left[\left\langle \left(S_{i}^{2}\right)^{2}\right\rangle -\left\langle S_{i}^{2}\right\rangle ^{2}\right]$
and gives the correct order of magnitude. There is also an additional
small $\Delta Z$ dependence, such that the ratio of the left- and
right-hand side varies by a factor of two over two orders of magnitude
change in $\Delta Z$. Presumably, this results from the plateau in
$B_{i}\left(S_{i}^{2}\right)$ at large $S_{i}^{2}$ in Fig.~1.

\textit{Distribution functions for $\Delta M_{i}$ and $S_{i}^{2}$:}
The distribution $P(B_{i})$, shown in Fig.~2(a), is also shown along
with $P(\mathcal{G}_{i})$ in Fig.~1 of Ref.~\cite{tuning_by_pruning}.
The distribution $P\left(\Delta B_{i}\right)$ is shown in Fig.~2(b),
and $P\left(\Delta\mathcal{G}_{i}\right)$ is shown in Fig.~2(c).
Compared to $P(B_{i})$ and $P(\mathcal{G}_{i})$, $P\left(\Delta B_{i}\right)$
and $P\left(\Delta\mathcal{G}_{i}\right)$ have a more robust power-law
scaling at small values: $P\left(\Delta M_{i}\right)\propto\left(\Delta M_{i}\right)^{-\kappa}$
with a common value of the exponent: $\kappa\sim0.5\pm0.03$. At large
values the distributions decrease with roughly exponential tails.
The most significant difference between $P\left(\Delta B_{i}\right)$
and $P\left(\Delta\mathcal{G}_{i}\right)$ is evident at large values,
where $P\left(\Delta B_{i}\right)$ develops a peak at large pressure
while $P\left(\Delta\mathcal{G}_{i}\right)$ continues to decay monotonically.
The distributions of $P\left(\Delta\mathcal{G}_{i}\right)$ are the
same in $d=2$ and $d=3$.

Note that as bonds are pruned, the distributions $P\left(\Delta B_{i}\right)$
and $P\left(\Delta\mathcal{G}_{i}\right)$ can evolve. We have therefore
studied the robustness of these distributions to the removal of a
small percentage of the bonds according to a variety of protocols.
We can prune bonds at random or we can prune according to the maximum
or minimum value of $\Delta M_{i}$. In all cases there is little
change in the functional form of the distribution $P\left(\Delta\mathcal{G}_{i}\right)$.
Remarkably, however, $P\left(\Delta B_{i}\right)$, which initially
differs from $P\left(\Delta\mathcal{G}_{i}\right)$, evolves to the
same distribution as $P\left(\Delta\mathcal{G}_{i}\right)$ in all
but the case where we took away bonds with the smallest values of
$\Delta B_{i}$. We only need to remove approximately\textcolor{green}{{}
}0.5\% of the bonds to achieve the collapse shown in Fig.~2(c). To
a good approximation, this distribution is given by $P\left(\Delta M\right)=\frac{1}{\sqrt{2\pi}}y^{-0.5}e^{-\frac{y}{2}}$
(see Fig.~2(c)).

To understand this universal distribution, we note that the bond angles
have only short-ranged correlations, the system is isotropic and $\mathcal{G}_{i}=\frac{1}{2Vk}\left(t_{\mathcal{G}i}\right)^{2}$,
where $t_{\mathcal{G}i}=k\sum_{j}e_{\mathcal{G}j}[S_{i}]_{j}$ is
the tension in a bond for a shear deformation. With these assumptions,
$t_{\mathcal{G}i}$ can be regarded as a sum of independent random
variables with zero mean. Since $[S_{i}]_{j}^{2}$ decays as function
of distance\cite{dipole_response,Jamming_SSS}, $t_{\mathcal{G}i}$
is dominated by the bonds that lie within this correlation length.
According to the central limit theorem, bonds with a given value of
$S_{i}^{2}$ are Gaussian distributed with zero average and a variance
$\left\langle \left(t_{\mathcal{G}i}\right)^{2}\right\rangle \propto\left\langle \mathcal{G}_{i}\left(S_{i}^{2}\right)\right\rangle \propto S_{i}^{2}$,
as shown in Fig 1. To place all bonds on the same scale we divide
the tension by $\sqrt{S_{i}^{2}}$, such that the overall distribution
of $t_{\mathcal{G}i}/\sqrt{S_{i}^{2}}$ is Gaussian. The distribution
of $\Delta\mathcal{G}_{i}\propto t_{\mathcal{G}i}^{2}/S_{i}^{2}$
can be then obtained through a change of variables. This leads to
\begin{equation}
P\left(\frac{\Delta\mathcal{G}_{i}}{\left\langle \Delta\mathcal{G}_{i}\right\rangle }=y\right)=\frac{1}{\sqrt{2\pi}}y^{-\frac{1}{2}}e^{-\frac{y}{2}},\label{eq:PDGi}
\end{equation}
consistent with our numerical results in Fig.~2(c).

Why does pruning bonds in many cases drive $\Delta B_{i}$ towards
the universal distribution of Eq.~\ref{eq:PDGi}? In contrast to
a shear deformation, the affine extension in compression $e_{Bi}=\epsilon\left|\delta r_{i}\right|$
is non-negative. As a result, we cannot assume that $t_{Bi}$ averages
to zero. Furthermore, at long distances $\left\langle [S_{i}]_{j}\right\rangle \propto\frac{1}{N}$;
this is a consequence of the special state of self stress that arises
just above the onset of jamming and accounts for the nonzero value
of the bulk modulus there. We suspect that pruning bonds destroys
this state of self stress so that once again, $t_{Bi}$ can be considered
the sum of uncorrelated random variables with zero average. In the
same manner, if $\left\langle B_{i}\right\rangle \propto S_{i}^{2}$,
then $t_{B_{i}}/\sqrt{S_{i}^{2}}$ have a zero average and are Gaussian
distributed. This leads directly to the universal distribution for
$P\left(\Delta B_{i}\right)$.

We also show the distribution $P(S_{i}^{2})$ in the solid curves
of Fig.~2(d). At low $S_{i}^{2}$: $P(S_{i}^{2})\sim\left(S_{i}^{2}\right)^{-\theta_{s}=-0.42\pm0.02}$.
This power law is robust to changes of $\Delta Z$. To understand
this, we argue that $P(S_{i}^{2})$ is directly related to the distribution
of interparticle forces, $P(F)$, in the original jammed system from
which the spring network was derived. Suppose the system has one bond
above the minimum needed for isostaticity, where there is only one
state of self stress. In this limit, force balance specifies a unique
set of forces on the bonds so that the state of self stress is uniquely
defined: $s_{i}\propto F_{i}$ \footnote{the proportionality constant is fixed by normalization $s_{i}$ to
have unit norm. } and $S_{i}^{2}=s_{i}^{2}$. The distribution of forces, $P(F)$ has
a power-law tail at small forces in this limit, such that $P(F)\sim F^{\theta=0.17462}$~\cite{Lerner2013,Charbonneau}.
Using a transformation of variables between $F_{i}$ and $S_{i}^{2}$
to obtain $P(S_{i}^{2})$ from $P(F_{i})$, we find $P\left(S_{i}^{2}\right)\propto\left(S_{i}^{2}\right){}^{(\theta-1)/2}$.
Thus, we predict $\theta_{s}=(1-\theta)/2=0.41269...$\cite{Charbonneau,Lerner2013},
in good agreement with the solid curves in Fig.~2(d). Note that the
result remains robust even as $\Delta Z$ increases well above the
minimum needed for rigidity. In the Supplemental Material we trace
this power-law to particles with the $d+1$ contacts \textendash{}
the minimum number of contacts needed for local stability.

\textit{Discussion.} At large length scales, periodic and disordered
networks are both governed by elastic theory and their macroscopic
mechanical response is captured by global elastic constants. At the
bond level, however, periodic and disordered networks exhibit completely
different behavior. For periodic networks, in which a unit cell of
a few nodes is repeated throughout, each bond $i$ has a similar local
modulus, $S_{i}^{2}$. In addition each bond plays a similar role
in resisting global deformations, so that $M_{i}$ is similar for
different global moduli $M$, and has a similar effect on those moduli
if it is removed. Thus $\Delta M_{i}$ is similar for different $M$.
Disordered networks are completely different\textendash the distributions
of $S_{i}^{2}$, $M_{i}$ and $\Delta M_{i}$ are broad and stretch
continuously down to zero. Variations in single-bond responses are
important not only for tuning global moduli, but also for controlling
the response of the system to stresses that are high enough to break
bonds, and ultimately to fracture the material.

We have shown that insight can be gained by studying a new local modulus
describing the response of a disordered network to the change of the
equilibrium length of bond $i$. This relates the contribution $M_{i}$
of a bond to a global modulus $M$, to the change of the modulus $\Delta M_{i}$
when bond $i$ is removed, 
and explains why $\mathcal{G}_{i}$ and $B_{i}$ have significant
correlations while $\Delta\mathcal{G}_{i}$ and $\Delta B_{i}$ do
not. We have further shown that the distribution of $M_{i}$ is universal
with a form that can be understood (at least after sufficient pruning).
With these results, we can now understand why the ratio of $\mathcal{G}/B$
is so tunable in disordered networks in terms of the local modulus
of a bond $i$. Tunability requires independence of bond-level response,
which relies on two properties: (1) that the distributions of $\Delta\mathcal{G}_{i}$
and $\Delta B_{i}$ are broad, continuous and extend continuously
to $\Delta\mathcal{G}_{i}=0$ and $\Delta B_{i}=0$, and (2) that
$\Delta\mathcal{G}_{i}$ and $\Delta B_{i}$ are uncorrelated. The
local modulus provides significant understanding of both of these
properties. 
\begin{acknowledgments}
We thank C. P. Goodrich, N. Pashine and J. P. Sethna for instructive
discussions. We acknowledge support from the US Department of Energy,
Office of Basic Energy Sciences, Division of Materials Sciences and
Engineering under Award DE-FG02-05ER46199 (AJL), the the Simons Foundation
for the collaboration ``Cracking the Glass Problem'' award \#348125
(DH),the Simons Foundation \#327939 (AJL), and the University of Chicago
MRSEC NSF DMR-1420709 (SRN). 
\end{acknowledgments}

\section{Supplementary material}

\subsection{Bond removal formula for non-identical spring constants}

In the main text, we derived the equation $\Delta M_{i}=M_{i}/S_{i}^{2}$
relating $\Delta M_{i}$, the change of the modulus $M$ when bond
$i$ is removed, to $M_{i}$, the contribution of bond $i$ to $M$,
and $S_{i}^{2}$, the local modulus, for the special case in which
all the bonds in the networks have the same spring constant. Here
we show that the same equation holds more generally, for arbitrary
spring constants on the bonds.

We follow the framework of Ref.~\cite{Lubensky_rev}, where the energy,
is given by :

\begin{equation}
E=\frac{1}{2}e_{s}^{T}\left(\left(k^{-1}\right)_{ss}\right)^{-1}e_{s}\label{eq:Energy}
\end{equation}
where $e_{s}$ is the projection of the affine bond extensions on
to the space of states of self stress (T denotes transpose); $k_{ij}=\delta_{ij}k_{i}$
is the matrix of the spring constants and the subscript $ss$ the
projection on to the space of state of self stress. We begin our analysis
by selecting an arbitrary basis of states of self stress $s_{\alpha}$
and rewrite Eq. 1 in this basis.

\begin{equation}
E=\frac{1}{2}\sum_{\alpha\beta}e_{\alpha}p_{\alpha\beta}^{-1}e_{\beta}
\end{equation}
where $e_{\alpha}=\sum_{i}e_{i}s_{\alpha,i}$. For convenience we
introduce $p_{\alpha\beta}=\left(k_{ss}^{-1}\right)_{\alpha\beta}=\sum_{ij}s_{\alpha,i}\left(k_{ss}^{-1}\right)_{ij}s_{\beta,j}$
which, by construction, is projected onto the space of states of self
stress.

Varying the spring constant of bond $i$ , modifies only a single
component in the spring constant matrix, $k_{ii}^{\prime-1}=k_{ii}^{-1}+\left(\frac{1}{k_{i}^{\prime}}-\frac{1}{k_{i}}\right)$.
The resulting change in $p$ is given by

\begin{equation}
p_{\alpha\beta}^{\prime}=p_{\alpha\beta}+\left(\frac{1}{k_{i}^{\prime}}-\frac{1}{k_{i}}\right)s_{\alpha,i}s_{\beta,i}
\end{equation}
To compute the resulting energy using Eq. 1 we require $\left(p'\right)^{-1}$
which can conveniently be computed using the Sherman\textendash Morrison
formula\cite{sherman1950},

\begin{equation}
p_{\alpha\beta}^{\prime-1}=p_{\alpha\beta}^{-1}-\frac{\left(\frac{1}{k'_{i}}-\frac{1}{k_{i}}\right)\sum_{\gamma\delta}p_{\alpha\gamma}^{-1}s_{\gamma,i}s_{\delta,i}p_{\delta\beta}^{-1}}{1+\left(\frac{1}{k'_{i}}-\frac{1}{k_{i}}\right)\sum_{\gamma\delta}s_{\gamma,i}p_{\gamma\delta}^{-1}s_{\delta,i}}.
\end{equation}
Removing a bond corresponds to taking the limit $k'_{i}\rightarrow0$,
which leads to

\begin{equation}
p_{\alpha\beta}^{\prime-1}=p_{\alpha\beta}^{-1}-\frac{\sum_{\gamma\delta}p_{\alpha\gamma}^{-1}s_{\gamma,i}s_{\delta,i}p_{\delta\beta}^{-1}}{\sum_{\gamma\delta}s_{\gamma,i}p_{\gamma\delta}^{-1}s_{\delta,i}}.
\end{equation}
Thus the change in energy is give by, 
\begin{equation}
\Delta E=\frac{1}{2}\frac{\sum_{\alpha\beta\gamma\delta}e_{\alpha}p_{\alpha\gamma}^{-1}s_{\gamma,i}s_{\delta,i}p_{\delta\beta}^{-1}e_{\beta}}{\sum_{\gamma\delta}s_{\gamma,i}p_{\gamma\delta}^{-1}s_{\delta,i}},\label{eq:ChangeEnergy}
\end{equation}
and all that remains is to recast this expression in terms of $M_{i}$
and $S_{i}^{2}$.

We begin with the denominator and show that it corresponds to a localized
deformation. To this end, we select $e_{j}=e_{0}\delta_{ij}$ in Eq.
\ref{eq:Energy} and find that $S_{i}^{2}$, defined as the local
modulus per-unit spring constant, is indeed given by the denominator.

\begin{equation}
S_{i}^{2}=\frac{2E}{k_{i}e_{0}^{2}}=\frac{1}{k_{i}}\sum_{\gamma\delta}s_{\gamma,i}p_{\gamma\delta}^{-1}s_{\delta,i}.
\end{equation}
We now turn to show that the numerator in Eq.~\ref{eq:ChangeEnergy}
is $M_{i}$. Following the derivation of Eq.~1, in Ref.~\cite{Lubensky_rev}
it is straightforward to show that the tension in a bond is given
by 
\begin{equation}
t=\left(\left(k^{-1}\right)_{ss}\right)^{-1}e_{s}.
\end{equation}
In our choice of basis, 
\begin{equation}
t_{i}=\sum_{\alpha\beta}s_{\alpha,i}p_{\alpha\beta}^{-1}e_{\beta}
\end{equation}
and therefore the numerator in Eq. \ref{eq:ChangeEnergy} is equal
to $t_{i}^{2}$.Recalling that the modulus $M=2E/e_{0}^{2}$, we find
that 
\begin{equation}
\Delta M_{i}=\frac{M_{i}}{S_{i}^{2}}.
\end{equation}

\subsection{Two-dimensional data}

In this section we test the robustness of the three dimensional results
presented in the paper, by comparing them to their two dimensional
counterparts. Fig. 1a and 1b shows $\left\langle B_{i}\left(S_{i}^{2}\right)\right\rangle $
and $\left\langle \mathcal{G}_{i}\left(S_{i}^{2}\right)\right\rangle $
for different values of $\Delta Z$. Similarly to the three dimensional
case, over a broad range these are proportional of $S_{i}^{2}$ with
$\left\langle \mathcal{G}_{i}\left(S_{i}^{2}\right)\right\rangle $
virtually independent of $\Delta Z$, while $\left\langle B_{i}\left(S_{i}^{2}\right)\right\rangle $
has a multiplicative dependence $\Delta Z^{\approx-1.4}$. Three dimensions
has a little different dependence on $\Delta Z$, $\left\langle B_{i}\left(S_{i}^{2}\right)\right\rangle \propto\Delta Z^{\approx-1}S_{i}^{2}$.
A possible source of this variation is that in two dimensions $\left\langle B_{i}\left(S_{i}^{2}\right)\right\rangle $
has a more pronounced plateau at large $S_{i}^{2}$ values.

Fig. 2 considers correlations between $B_{i}$ and $\mathcal{G}_{i}$,
and the correlations between $\Delta B_{i}$ and $\Delta\mathcal{G}_{i}$.
Fig. 2a shows that distribution of $B_{i}$ depends on the range $\mathcal{G}_{i}$
values implying that these are correlated. On the other hand the distribution
of $\Delta B_{i}$, shown in Fig. 2b, appears to be independent of
$\Delta\mathcal{G}_{i}$ suggesting tiny amount of correlations. The
behavior in two dimensions appear identical to the behavior in three
dimensions. We also find little difference in the distributions of
$B_{i}$, $\mathcal{G}_{i}$, $\Delta B_{i}$ and $\Delta\mathcal{G}_{i}$
between two and three dimensions and in fact, Fig. 2c in main text
shows cases where they are identical.

Fig. 3 shows the distribution of $S_{i}^{2}$ as a function of $\Delta Z$
. Also here, there is no apparent difference from the three dimensional
case and the exponent characterizing the power-law scaling at small
$S_{i}^{2}$ agrees with the prediction $P\left(S_{i}^{2}\right)\propto\left(S_{i}^{2}\right)^{-0.41269}$.

\begin{figure}[H]
\begin{centering}
\includegraphics[scale=0.35]{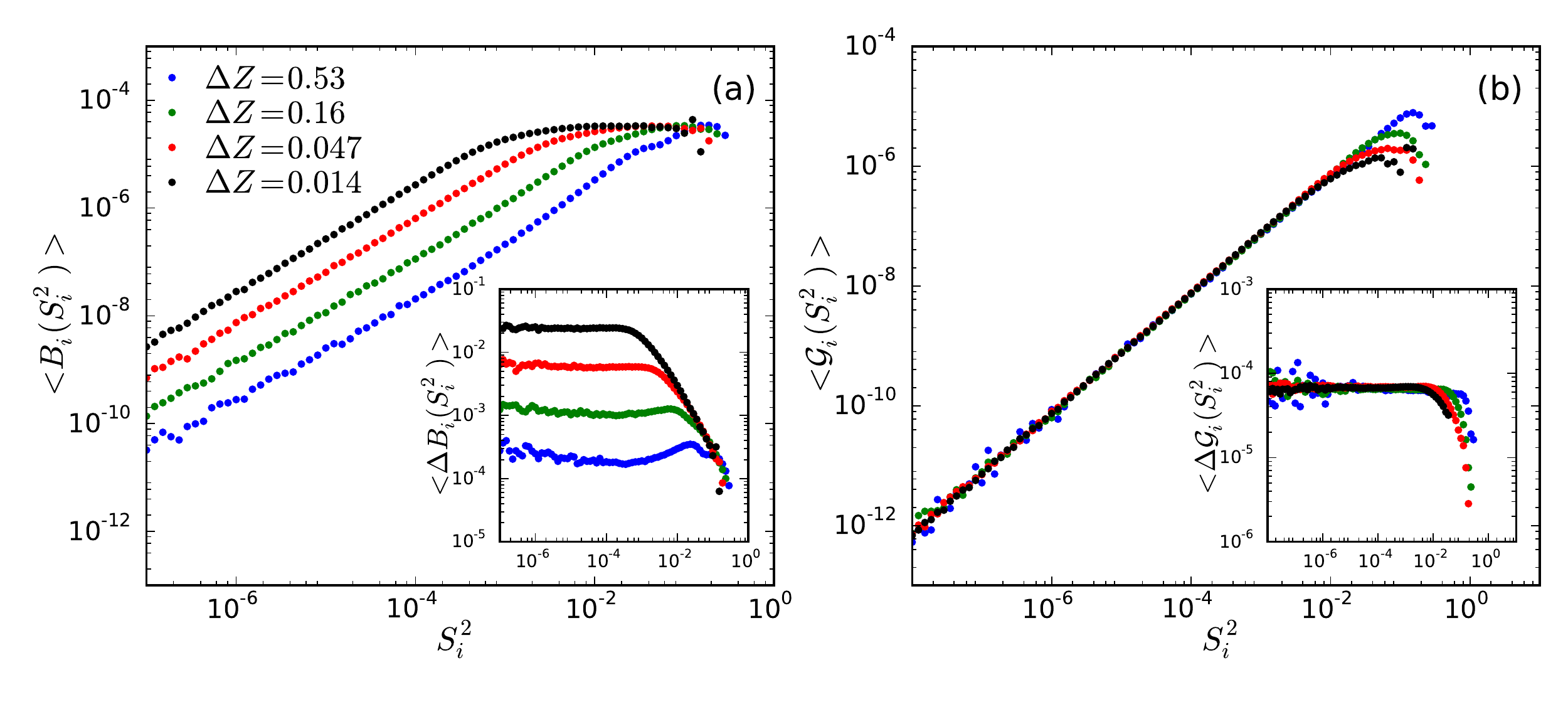} 
\par\end{centering}
\caption{a) The conditional average $\left\langle B_{i}\left(S_{i}^{2}\right)\right\rangle $
and b) $\left\langle \mathcal{G}_{i}\left(S_{i}^{2}\right)\right\rangle $
are proportional to $S_{i}^{2}$ over a broad range of $S_{i}^{2}$.
In the inset the corresponding $\left\langle \Delta B_{i}\left(S_{i}^{2}\right)\right\rangle $
and $\left\langle \Delta\mathcal{G}_{i}\left(S_{i}^{2}\right)\right\rangle $
are shown. To good approximation these are independent of $S_{i}^{2}$
except at high $S_{i}^{2}$. Data is from 2d systems with $N=8192$
particles.}
\end{figure}

\begin{figure}[H]
\begin{centering}
\includegraphics[scale=0.35]{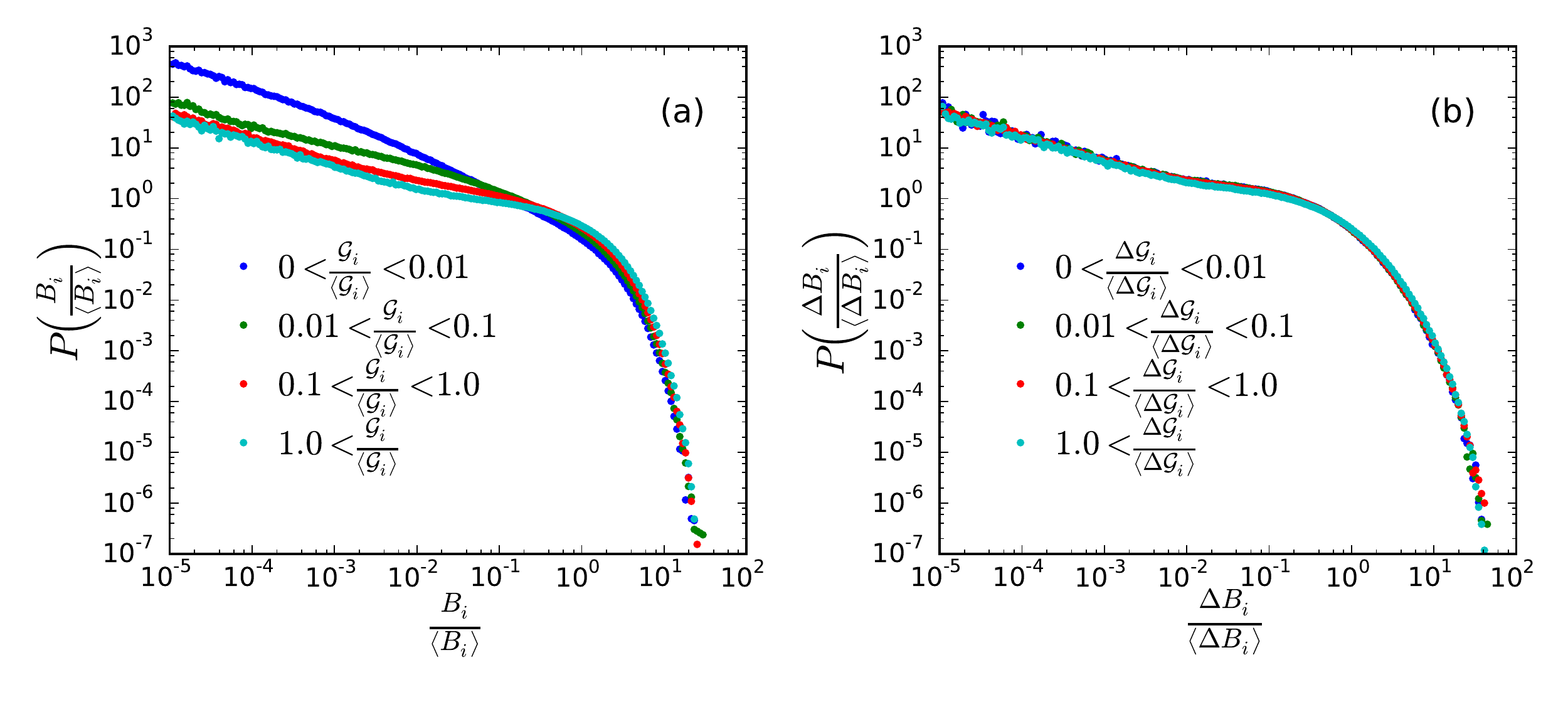} 
\par\end{centering}
\caption{a) $P\left(B_{i}\right)$ for different ranges of $\mathcal{G}_{i}$
and b) $P\left(\Delta B_{i}\right)$ for different range of $\Delta\mathcal{G}_{i}$
in two dimensions. The collapse indicates almost vanishing correlations.
Here $\Delta Z=0.047$ and the number of particles $N=8192$.}
\end{figure}

\begin{figure}[H]
\begin{centering}
\includegraphics[scale=0.4]{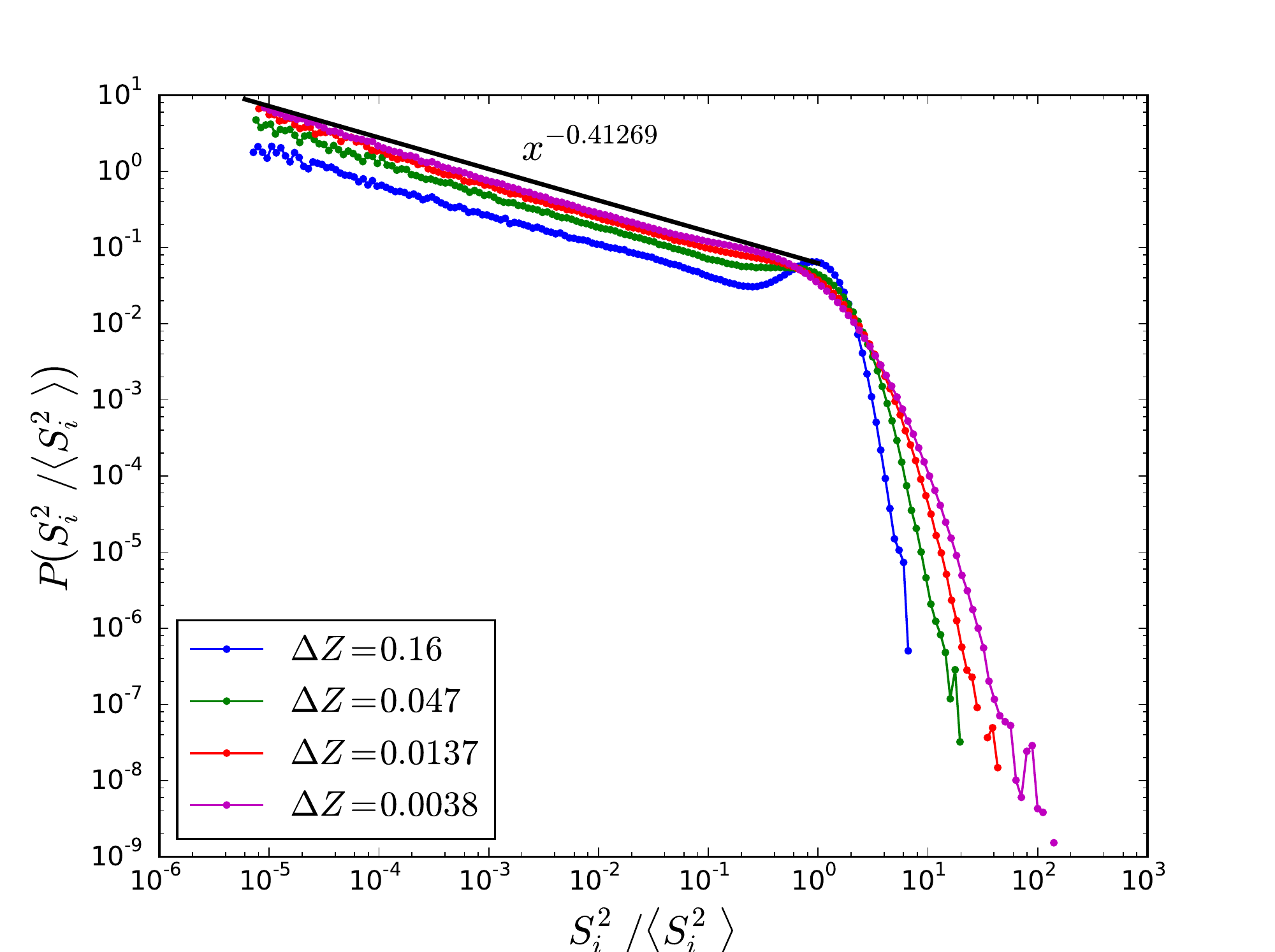} 
\par\end{centering}
\caption{$P\left(S_{i}^{2}\right)$ at different values of $\Delta Z$ in two
dimensions with $N=8192$ particles. The exponent $0.41269...$ is
the prediction based on the contribution from bucklers\cite{Charbonneau,Lerner2013}.
Note that as $\Delta Z$ is increased a peak in $P\left(S_{i}^{2}\right)$
develops. }
\end{figure}

\subsection{Effect of ``bucklers\textquotedbl{} on $P\left(S_{i}^{2}\right)$}

In this section we show that the scaling of $P\left(S_{i}^{2}\right)\propto\left(S_{i}^{2}\right)^{-\theta_{S}\approx-0.42}$
at small $S_{i}^{2}$ results from particular particles with $d+1$
neighbors called ``bucklers''. To this end we consider the distribution
of $S_{i}^{2}$ when these particles are not included. The scaling
argument in the main text is based on the behavior at isostaticity
where the distribution of forces, $P\left(F\right)\propto F^{\theta}$
at small $F$. As we argued in the main text 
\begin{equation}
\theta_{s}=1/2-\theta/2.\label{eq:thetaeq}
\end{equation}
The exponent $\theta$ has two contributions~\cite{Charbonneau}
\textendash{} (1) The mean-field exponent \cite{mean_field1,mean_field2}
$\theta^{\left(\infty\right)}=0.42311...$ and (2) The exponent due
to ``buckler'' particles $\theta=0.17462...$, which overshadows
the first contribution. Buckler particles are those with $d+1$ interacting
neighbors in $d$ dimensions, for which $d$ forces are nearly balanced
across the particle in what is nearly a line in $d=2$ or a plane
in $d=3$, while the remaining force is very small. In the main text,
we showed that Eq.~\ref{eq:thetaeq} holds if all particles and forces
are included. If bucklers are removed, the force distribution scales
as $P(F)\sim F^{\theta^{(\infty)}}$ at small $F$ \cite{Charbonneau}.
We would therefore expect the exponent in $P(S_{i}^{2})$ to change
when bucklers are removed. Indeed, the prediction of Eq.~\ref{eq:thetaeq}
that $\theta_{S}^{(\infty)}=0.28845...$ is in good agreement with
our numerical results at the lowest value of $\Delta Z$ shown in
Fig. 4. Note that once bucklers are removed, however, the exponent
$\theta_{S}$ is not robust to changes in $\Delta Z$; Fig.~4 shows
that $P\left(S_{i}^{2}\right)$ approaches a constant at small $S_{i}^{2}$
as $\Delta Z$ increases. Comparing Fig.~4 to Fig.~2d of the main
text we deduce that bucklers are the origin of the small $P\left(S_{i}^{2}\right)$
scaling seen in Fig.~2d of the main text, which, interestingly, depends
only weakly on $\Delta Z$.

\begin{figure}[H]
\begin{centering}
\includegraphics[scale=0.35]{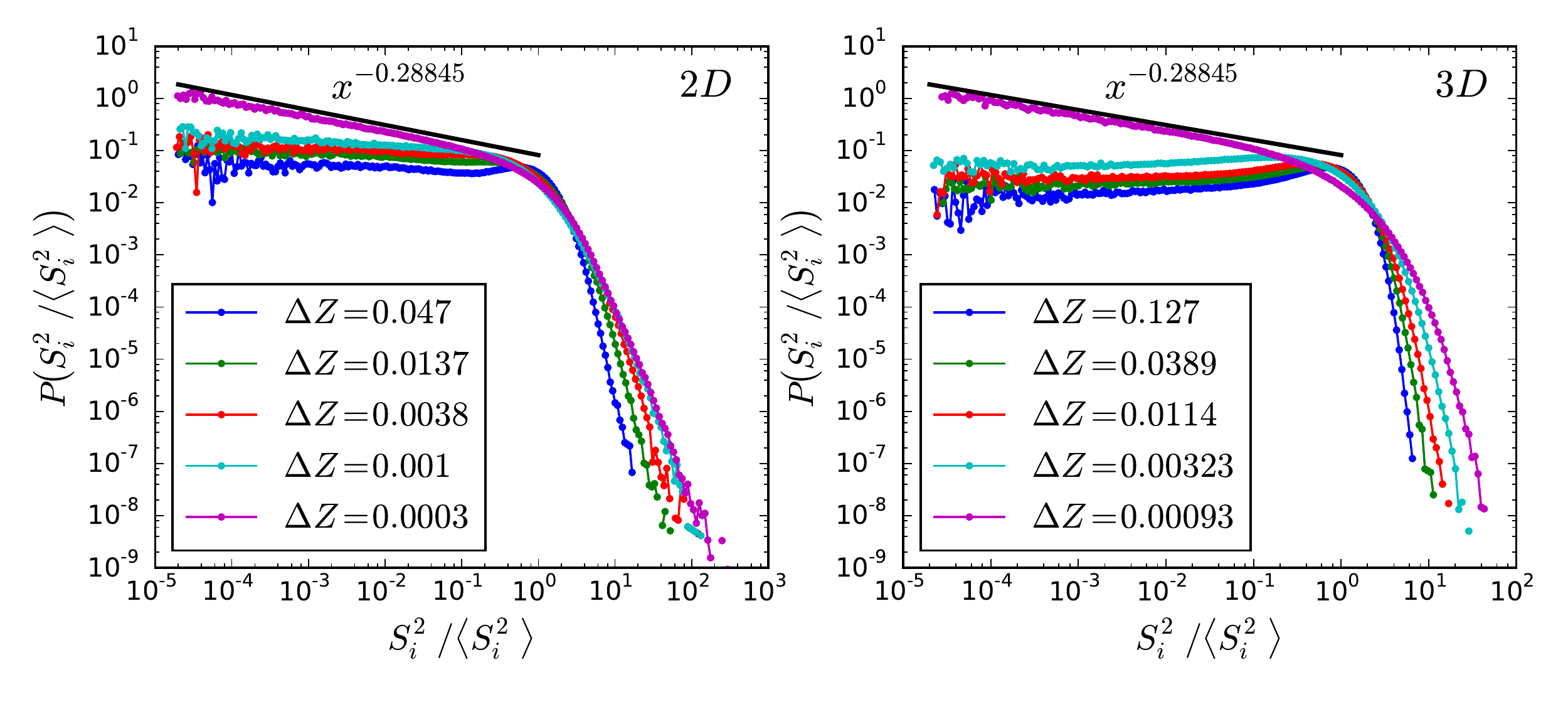} 
\par\end{centering}
\caption{$P\left(S_{i}^{2}\right)$ in two dimensions (left) and three dimensions
(right) when buckler particles are removed. The exponent $0.28845$
is the prediction based on the the mean-field scaling of $P\left(F\right)$
at isostaticity . Note that the smallest $\Delta Z$ curve is different
from the remaining curves with larger $\Delta Z$, suggesting that
the exponent is not robust to the increase of $\Delta Z$ unlike the
buckler contribution. }
\end{figure}

 \bibliographystyle{apsrev4-1}
\bibliography{biblo}

\end{document}